\documentclass{article}
\usepackage{amssymb}
\usepackage{cite}
\usepackage{ifpdf}

\def\R{{\mathbb{R}}}

\def\C{\mathbb{C}}

\def\H{{\mathcal{H}}}
\date{\empty}
\def\<{\langle}
\def\>{\rangle}
\def\tr{{\rm{tr}}}
\begin{document}
\title{\bf Some remarks on Bell non-locality and Einstein-Podolsky-Rosen steering of bipartite states}
\author{Huaixin Cao, Zhihua Guo\\
{\it\small School of Mathematics and Information Science, Shaanxi Normal University}\\
{\it\small Xi'an 710119, China}\\
{\it\small Email: caohx@snnu.edu.cn, guozhihua@snnu.edu.cn}} \vskip
0.1in
\maketitle

{\bf Abstract}. Bell nonlocality and Einstein-Podolsky-Rosen (EPR) steering are every important quantum correlations of a composite quantum system. Bell nonlocality of a bipartite state is a quantum correlation demonstrated by some local quantum measurements, while EPR steering is another form of quantum correlations, observed firstly by Schrodinger in the context of famous EPR paradox. In this paper, we give some remarks on Bell nonlocality and EPR steering of bipartite states, including  mathematical definitions and characterizations of these two quantum correlations, the convexity and closedness of the set of all Bell local states and the set of all EPR unsteerable states. We also derive a EPR-steering criteria, with which the EPR steerability of the maximally entangled states are checked.


\section{Introduction}

Generally, quantum correlations means the correlations between subsystems of a composite quantum system, including Bell nonlocality, steerability, entanglement and quantum discord.

Bell nonlocality of a bipartite state is a  quantum correlation demonstrated by some local quantum measurements whose statistics of the measurement outcomes cannot be explained by a local hidden variable (LHV) model \cite{Bell,BCPSW}. Such
a nonclassical feature of quantum mechanics can be used in
device-independent quantum information processing \cite{BCPSW}. For more works
on Bell nonlocality, please refer to Clauser and Shimony \cite{Clauser}, Home and
Selleri \cite{Home}, Khalfin and Tsirelson \cite{Khal},
Tsirelson \cite{Tsire}, Zeilinger \cite{Zeilinger}, Werner and Wolf \cite{Werner},
Genovese \cite{Genovese}, and Buhrman et al. \cite{Buhrman}, and references
therein.

Einstein-Podolsky-Rosen (EPR) steering  as a form of quantum correlations, was first observed by Schrodinger  \cite{Schro} in the context of famous
Einstein-Podolsky-Rosen (EPR) paradox \cite{Einstein,paradox1,paradox2,paradox3}. EPR steering arises in the scenario where some local quantum measurements on
one part of a bipartite system are used to steer the other part. This scenario demonstrates EPR steering if the obtained ensembles cannot be explained by a local hidden
state (LHS) model \cite{Wiseman}. Followed in close analogy with criteria for other forms of quantum nonlocality (Bell nonlocality and entanglement), Cavalcanti et al. \cite{Cava} developed a general theory of experimental EPR-steering criteria and derived a number of criteria applicable to discrete as well as continuous-variable observables. Saunders et al. \cite{Saun} contributed experimental EPR-steering by using Bell local states. Bennet et al. \cite{Bennet} derived arbitrarily loss-tolerant tests, which enable us to perform a detection-loophole-free demonstration of Einstein-Podolsky-Rosen steering with parties separated by a coiled 1-km-long optical fiber.
H\"{a}ndchen et al. \cite{Eberle} presented
an experimental realization of two entangled Gaussian modes of
light that in fact shows the steering effect in one direction but
not in the other. The generated one-way steering gives a new
insight into quantum physics and may open a new field of
applications in quantum information.

EPR steering, as a form of bipartite quantum correlation that is intermediate between entanglement and Bell nonlocality, allows for entanglement certification when the measurements performed by one of the parties are not characterized (or are untrusted) and has applications in quantum key
distribution.   Branciard et al. \cite{Bran}  analyzed the security and feasibility of a protocol for quantum key distribution (QKD) in a context where only one of the two parties trusts his measurement apparatus and clarified the link between the security of this one-sided DI-QKD scenario and the demonstration of quantum steering, in analogy to the link between DI-QKD and the violation of Bell inequalities.
Wittmann et al. \cite{Witt}  presented the first loophole-free demonstration of EPR-steering by violating three-setting quadratic steering inequality in light of polarization entangled photons shared between two distant laboratories.
Steinlechner et al.  \cite{Stein} achieved  an unprecedented low conditional variance product of about $0.04<1$, where $1$ is the upper bound below which steering is present, and observed the steering effect on an unconditional two-mode-squeezed entangled state that contained a total vacuum state contribution of less than $8\%$.
Reid \cite{Reid} proved that EPR paradox can be used to verify that the quantum benchmark for qubit teleportation has been reached, without postselection and  EPR steering inequalities involving
$m$ measurement settings can also be used to confirm quantum teleportation if one assumes trusted detectors for Charlie and Alice.  Skrzypczyk et al. \cite{Skrzy}  proposed a way of quantifying this phenomenon and use it to study the steerability of several quantum states and shown that every pure entangled state is maximally steerable and the projector onto the antisymmetric subspace is maximally steerable for all dimensions.

Piani et al. \cite{NecSuf}   provided a necessary and sufficient characterization of steering, based on a quantum information processing task: the discrimination of branches in a quantum evolution, which we dub subchannel discrimination. They also proved that, for any bipartite steerable state, there are instances of the quantum subchannel discrimination problem for which this state allows a correct discrimination with strictly higher probability than in absence of entanglement, even when measurements are restricted to local measurements aided by one-way communication. Many of the standard Bell inequalities (e.g. CHSH ) are not effective for detection of quantum correlations which allow for steering, because for a wide range of such correlations they are not violated. Zukowski et al. \cite{Zuko} presented some  Bell like inequalities which have lower bounds for non-steering correlations than for local causal models. These inequalities involve all possible measurement settings at each side.
Geometric Bell like inequalities for steering.

By definition, it is easy to check that every separable state is unsteerable state and any unsteerable state is Bell local. Thus, quantum states that demonstrate
Bell nonlocality form a subset of EPR steerable states which, in turn, form a subset of entangled states. Furthermore, Quintino et al. proved in \cite{Inequi} that entanglement, one-way steering, two-way steering, and Bell nonlocality are genuinely
different. Specifically, considering general POV measurements,
they proved the existence of (i) entangled states
that cannot lead to steering, (ii) states that can lead to
steering but not to Bell nonlocality, and (iii) states which
are one-way steerable but not two-way steerable.

Zhu et al. \cite{Zhu} proposed a general framework for constructing universal steering criteria that are applicable to arbitrary bipartite states and measurement settings of the steering party. The same framework is also useful for studying the joint measurement problem. Based on the data-processing inequality for an extended R\'{e}nyi relative entropy, they also introduced a family of steering inequalities, which detect steering much more
efficiently than those inequalities known before. Sun et al. \cite{Sun}	experimentally demonstrated asymmetric EPR steering for a class of two-qubit states in the case of two measurement settings and proposed a practical method to quantify the steerability. They also provided a necessary and sufficient condition for EPR steering and clearly demonstrate one-way EPR steering.

Recently, Cavalcanti et al. \cite{CP} contributed    a review on quantum steering with focus on semidefinite programming. Moreover, based on decomposing the measurement correlations in terms of extremal boxes of the steering scenario, Das et al. \cite{JM} presented a method to check EPR steering in the scenario where the steering party performs two black-box measurements and the trusted party performs two mutually unbiased projective qubit measurements. In this context, they proposed a measure of
steerability called steering cost and proved that their steering cost is a convex steering monotone.

In this paper, we will give some remarks on Bell nonlocality and EPR steering of bipartite states, including  mathematical definitions and characterizations of these two quantum correlations, the convexity and closedness of the set of all Bell local states and the set of all EPR unsteerable states. We also derive a EPR-steering criteria, with which the EPR steerability of the maximally entangled states are checked. The other parts of this note are divided as follows. In Section 2, we will give the definition of Bell locality and Bell nonlocality of bipartite states, and establish some equivalent characterizations of Bell locality. Moreover, we will prove that the closedness and convexity of the set of all Bell local states. In Section 3, we will give the definitions of PER unsteerability and PER steerability of bipartite states, and establish some equivalent characterizations of PER unsteerability. Moreover, we will prove that the closedness and convexity of the set of all unsteerable states. In Section 4, we will establish a EPR steering criteria and prove the EPR steerability of the maximally entangled states.

\section{Bell nonlocality}

In what follows, we use $\mathcal{H}_A$ and $\mathcal{H}_A$ to denote two finite dimensional   complex Hilbert spaces, which describe two quantum systems $A$ and $B$, respectively. We also use $\mathcal{D}_X$ to denote the set $D(\mathcal{H}_X)$ of all quantum states of the system $X$ described by a Hilbert space $\mathcal{H}_X$.

A standard nonlocality scenario (SNLS) consists of two distant
systems on which two observers, Alice and Bob, perform
respectively $m_A$ and $m_B$ different measurements of $o_A$ and $o_B$
possible outcomes. More explicitly, when the outcomes of Alice and Bob are
labeled $a$ and $b$,  respectively,  while their POV measurement choices are
$$M^x=\{M_{a|x}:a = 1,\ldots,o_A\}(x = 1,\ldots,m_A),$$$$N^y=\{N_{b|y}:b = 1,\ldots,o_B\}(
y = 1,\ldots,m_B),$$  respectively, the family
$$\mathcal{M}_{AB}\equiv\mathcal{M}_A\otimes\mathcal{N}_B:=\{M^x\otimes N^y: x = 1,\ldots,m_A,y = 1,\ldots,m_B\}$$
is said to be a {\it standard nonlocality scenario (SNLS)} for system $AB$,
where
$$\mathcal{M}_A=\{M^x:x=1,2,\ldots,m_B\}, \ \mathcal{N}_B=\{N^y:y=1,2,\ldots,m_B\},$$ called {\it measurement assemblages} of $A$ and $B$, respectively, and
$$M^x\otimes N^y=\{M_{a|x}\otimes N_{b|y}: a = 1,\ldots,o_A, b = 1,\ldots,o_B\}.$$

{\bf Definition 2.1.}  Let $\rho^{AB}$ be a state of the system $AB$, $\mathcal{M}_A=\{M^x\}_{x=1}^{m_A}$ and $\mathcal{N}_B=\{N^y\}_{y=1}^{m_B}$  be two sets of some POV measurements (POVMs) of $A$ and $B$, respectively.

 (1) A state $\rho^{AB}$ is said to be {{Bell local}} for
$\mathcal{M}_A\otimes\mathcal{N}_B$ if there exist a probability  distribution (PD)
 $\{\pi_\lambda\}_{\lambda=1}^d$ such that for each $(\lambda,x)$ and each $(\lambda,y)$, there exist PDs  $\{P_A({a|x},\lambda)\}_{a=1}^{o_A}$ and $\{P_B({b|y},\lambda)\}_{b=1}^{o_B}$, respectively,  such that
$${\rm{tr}}[(M_{a|x}\otimes N_{b|y})\rho^{AB}]=
\sum_{\lambda=1}^d\pi_\lambda P_A({a|x},\lambda)P_B({b|y},\lambda),\ \forall a,b,x,y.\eqno(2.1)$$

In this case, Eq. (2.1) is said to be a {\it local hidden variable (LHV) model} of  $\rho^{AB}$  with respect to $\mathcal{M}_{AB}$ and $\lambda$ is said to be a {\it local hidden variable.}
Denote by $\mathcal{BL}(\mathcal{M}_A, \mathcal{N}_B)$ the set of all states $\rho^{AB}$ that are Bell local for $\mathcal{M}_A\otimes\mathcal{N}_B$.

(2) A state $\rho^{AB}$ is said to be {{\it Bell nonlocal}}  for   $\mathcal{M}_A\otimes\mathcal{N}_B$ if it is not Bell local for   $\mathcal{M}_A\otimes\mathcal{N}_B$.
Denote  by $\mathcal{BNL}(\mathcal{M}_A, \mathcal{N}_B)$ the set of all states $\rho^{AB}$ that are Bell nonlocal for $\mathcal{M}_A\otimes\mathcal{N}_B$.

(3) A state $\rho^{AB}$ is said to be {{\it Bell local}} if  for every $\mathcal{M}_A\otimes\mathcal{N}_B$, there exists a PD
 $\{\pi_\lambda\}_{\lambda=1}^d$ such that Eq. (2.1) holds.
Denote by $\mathcal{BL}(AB)$ the set of all Bell local states $\rho^{AB}$ of $AB$.

(4) A state $\rho^{AB}$ is said to be {{\it Bell nonlocal}} if it is not Bell local, i.e. there exists an $\mathcal{M}_A\otimes\mathcal{N}_B$ such that $\rho^{AB}$ is not Bell local for  $\mathcal{M}_A\otimes\mathcal{N}_B$.
Denote  by $\mathcal{BNL}(AB)$ the set of all states $\rho^{AB}$ that are Bell nonlocal.

{\bf Remark 2.1.}  By definition above, we see that when a state $\rho^{AB}$ is  {{Bell local}} for
$\mathcal{M}_A\otimes\mathcal{N}_B$, it has an LHV model (2.1).
Finding the sums of two sides for $b=1,2,\ldots,o_b$ yields that
$${\rm{tr}}[(M_{a|x}\otimes I_B)\rho^{AB}]=
\sum_{\lambda=1}^d\pi_\lambda P_A({a|x},\lambda),\ \forall a,x.$$
This shows that the measurement results of Alice with $\mathcal{M}_A$ are independent of the measurements of Bob. Similarly, we have
$${\rm{tr}}[(I_A\otimes N_{b|y})\rho^{AB}]=
\sum_{\lambda=1}^d\pi_\lambda P_B({b|y},\lambda),\ \forall b,y,$$
implying that the measurement results of Bob with $\mathcal{N}_B$ are independent of the measurements of Alice. Moreover, we see from definition that

{\bf Bell local states:} $\mathcal{BL}(AB)=\bigcap_{\mathcal{M}_A, \mathcal{N}_B}\mathcal{BL}(\mathcal{M}_A, \mathcal{N}_B);$

{\bf Bell nonlocal states:}
$\mathcal{BNL}(AB)=\bigcup_{\mathcal{M}_A, \mathcal{N}_B}\mathcal{BNL}(\mathcal{M}_A, \mathcal{N}_B).$

By Definition 2.1, we know that

{\bf Remark 2.2.}  Every separable state is Bell local. Equivalently, Bell nonlocal state must be entanglement.

To see this, let $\rho^{AB}=\sum_{\lambda=1}^dc_\lambda\rho^A_\lambda\otimes\rho^B_\lambda$ be separable. Then for every  $\mathcal{M}_A\otimes \mathcal{N}_B$, we have
$${\rm{tr}}[(M_{a|x}\otimes N_{b|y})\rho^{AB}]=\sum_{\lambda=1}^d\pi_\lambda P_A(a|x,\lambda)P_B(b|y,\lambda),\ \ \forall a,b,x,y,$$
where
$$P_A(a|x,\lambda)={\rm{tr}}(M_{a|x}\rho^A_\lambda), \ \ P_B(b|y,\lambda)={\rm{tr}}(N_{b|y}\rho^B_\lambda).$$
By Definition 2.1, $\rho^{AB}$ is Bell local. Note that in this case, response functions $P_A(a|x,\lambda)$ and $P_B(b|y,\lambda)$ are ``quantum", i.e. they are induced by quantum states.

\textbf{\bf Remark 2.3.} In the definition of locality of a state, the probability  distribution $\{\pi_\lambda\}_{\lambda=1}^d$ of the hidden variable $\lambda$ is necessary. Generally, the dimension $d$ of hidden variable space depends on not only the measurement assemblage $\mathcal{M}_A\otimes \mathcal{N}_B$ but also the state $\rho^{AB}$.

An expectation is to find the same dimension of hidden variable spaces for all Bell local states for a given $\mathcal{M}_A\otimes \mathcal{N}_B$.
To do this, let us consider the set $\Omega_A$ of all possible maps from $S_m=\{1,2,\ldots,m_A\}$ into $S_o=\{1,2,\ldots,o_A\}$. Clearly, $\Omega_A$ has just $N_A:=o_A^{m_A}$ elements and so can be written as
 $$\Omega_A=\{J_1,J_2,\ldots,J_{N_A}\}.$$
Each element $J$ of $\Omega$ denotes a ``measurement scenario", which assigns an outcome value $a$ for each POVM $\mathcal{M}^x$, that is, $J(x)=a$. We use $p_A(k,\lambda)$ to denote the probability of a measurement scenario $J_k$ to be used when Alice receives a classical message $\lambda$ in $\Lambda$. Thus, $\{p_A(k,\lambda)\}_{k=1}^{N_A}$ is a PD and depending only on the number $m_A$ of measurement operators and the number $o_A$ of the common  outcomes. Let $P(a,x,\lambda)$ be the probability of obtaining the
outcome $a$ when Alice receives a classical message $\lambda$ in $\Lambda$ and uses $\mathcal{M}^x$. Then the total probability formula yields that
$$P(a,x,\lambda)=\sum_{k=1}^{N_A}p_A(k,\lambda)\delta_{a,J_k(x)},\ \ \forall a\in S_o.\eqno(2.2)$$

Similarly, let $P(b,y,\lambda)$ be the probability of obtaining the
outcome $b$ when Bob receives a classical message $\lambda$ in $\Lambda$, and
$\Omega_B$ the set
 of all possible maps from $T_m=\{1,2,\ldots,m_B\}$ into $T_o=\{1,2,\ldots,o_B\}$. Clearly, $\Omega_B$ has just $N_B:=o_B^{m_B}$ elements and so can be written as
 $$\Omega_B=\{K_1,K_2,\ldots,K_{N_B}\}.$$
Then
$$P(b,y,\lambda)=\sum_{j=1}^{N_B}p_B(j,\lambda)\delta_{a,K_j(y)},\ \ \forall b\in T_o,\eqno(2.3)$$
where $N_B:=o_B^{m_B}$ is the number of elements $K_j$'s of $\Omega_B$.

When a state $\rho^{AB}$ is {{Bell local}}  for   $\mathcal{M}_A\otimes \mathcal{N}_B$, it  has an LHV model (2.1). Thus, for every $k$ we have
$$\sum_{a}P_A(a|x,k)=1(1\le x\le m_A),\ \sum_{b}P_B(b|y,k)=1(1\le y\le m_B).$$
By finding the sums of two sides of
(2.1) for $b\in T_o$, we get that
$${\rm{tr}}[(M_{a|x}\otimes I_B)\rho^{AB}]=
\sum_{\lambda=1}^d\pi_\lambda P_A({a|x},\lambda),\ \forall a,x.\eqno(2.4)$$
Likewise,
$${\rm{tr}}[(I_A\otimes N_{b|y})\rho^{AB}]=
\sum_{\lambda=1}^d\pi_\lambda P_B({b|y},\lambda),\ \forall b,y.\eqno(2.5)$$
The left-hand side of (2.4) is the probability of obtaining outcome $a$ when the measurement $\mathcal{M}^x$ is used. The quantity $\pi_\lambda$ can be viewed as the probability of Alice receiving a message $\lambda$, and  the quantity $P_A({a|x},\lambda)$ should be the probability of obtaining outcome $a$ when the measurement $\mathcal{M}^x$ is used and a message $\lambda$ is received by Alice. From Eqs. (2.2) and (2.3), we know that
$$P_A({a|x},\lambda)=\sum_{k=1}^{N_A}p_A(k,\lambda)\delta_{a,J_k(x)},\ \ \forall a\in S_o,\eqno(2.6)$$
where $\sum_{k=1}^{N_A}p_A(k,\lambda)=1$ for all $\lambda$.
Similarly,
$$P_B({b|y},\lambda)=\sum_{j=1}^{N_B}p_B(j,\lambda)\delta_{b,K_j(y)},\ \ \forall b\in T_o,\eqno(2.7)$$
where $\sum_{j=1}^{N_B}p_B(j,\lambda)=1$   for all $\lambda$. It follows from (2.6), (2.7) and (2.1) that
$${\rm{tr}}[(M_{a|x}\otimes N_{b|y})\rho^{AB}]=\sum_{k=1}^{N_A}\sum_{j=1}^{N_B}q_{k,j}\delta_{a,J_k(x)}\delta_{b,K_j(y)},\eqno(2.8)$$
where $q_{k,j}=\sum_{\lambda=1}^d\pi_\lambda p_A(k,\lambda)p_B(j,\lambda)\ge0$ for all $k,j$ satisfying
 $\sum_{k=1}^{N_A}\sum_{j=1}^{N_B}q_{k,j}=1.$

Conversely, if there exist a probability distribution
$$\{q_{k,j}:1\le k\le N_A,1\le j\le N_B\}:=\{\pi_1,\pi_2,\ldots,\pi_{N_AN_B}\}$$ satisfying (2.8), then Eq. (2.1) holds for
$$P_A(a|x,\lambda)=\delta_{a,J_k(x)}\mbox{\ and\ } P_B(b|y)=\delta_{b,K_j(y)}\mbox{\ if\ } \pi_\lambda=q_{k,j}$$
and then a state $\rho^{AB}$ is Bell local  for   $\mathcal{M}_A$ and $\mathcal{M}_B$.

As a result, we have the following conclusion.

{\bf Theorem 2.1.} {\it A state $\rho^{AB}$ is {{Bell local}}  for  $\mathcal{M}_A\otimes\mathcal{N}_B$ if and only if there exists a probability distribution $\{q_{k,j}:1\le k\le N_A,1\le j\le N_B\}$ satisfying Eq. (2.8).}

This characterization of Bell locality is very useful due to the sum in (2.8) was taken for a fixed number $N_AN_B$ of terms, the PDs $\{\delta_{a,J_k(x)}\}_{a=1}^{o_A}$ depending only on $M_{a|x}$ and $\{\delta_{b,K_j(y)}\}_{b=1}^{o_B}$  depending only on $N_{b|y}$ are independent of $\rho^{AB}$, while the PD $\{q_{k,j}:1\le k\le N_A,1\le j\le N_B\}$ depends only on $\rho^{AB}$. For instance, we can prove the following conclusion by using this characterization.

\textbf{\bf Corollary 2.1.} {\it The set $\mathcal{BL}(\mathcal{M}_A,\mathcal{N}_B)$  is a compact convex subset of $\mathcal{D}_{AB}$. Furthermore, $\mathcal{BL}(AB)$ is a compact  convex set.}

{\bf Proof.} Let $\{\rho_n\}_{n=1}^\infty\subset \mathcal{BL}(\mathcal{M}_A,\mathcal{N}_B)$ with $\rho_n\rightarrow\rho$ as $n\rightarrow\infty.$ We see from Theorem 2.1 that for each $n$, there exists a PD $\{q^n_{k,j}:1\le k\le N_A,1\le j\le N_B\}$ such that
$${\rm{tr}}[(M_{a|x}\otimes N_{b|y})\rho_n]
=\sum_{k=1}^{N_A}\sum_{j=1}^{N_B}q^n_{k,j}\delta_{a,J_k(x)}\delta_{b,K_j(y)},\ \forall a,x,b,y,\eqno(2.9)$$
for $n=1,2,\ldots.$ By choosing subsequence, we may assume that for each $(k,j)$, the sequence $\{q^n_{k,j}\}_{n=1}^\infty$ is convergent, sat $q^n_{k,j}\rightarrow q_{k,j}$ as  $n\rightarrow\infty$. Clearly,  $\{q_{k,j}:1\le k\le N_A,1\le j\le N_B\}$ is a PD.
Letting $n\rightarrow\infty$ in Eq. (2.9) yields that
$${\rm{tr}}[(M_{a|x}\otimes N_{b|y})\rho]
=\sum_{k=1}^{N_A}\sum_{j=1}^{N_B}q_{k,j}\delta_{a,J_k(x)}\delta_{b,K_j(y)},\ \forall a,x,b,y.$$
By Theorem 2.1, we conclude that $\rho\in\mathcal{BL}(\mathcal{M}_A,\mathcal{N}_B)$. This shows that $\mathcal{BL}(\mathcal{M}_A,\mathcal{N}_B)$ is closed and then compact due to the compactness of $\mathcal{D}_{AB}.$

To check the convexity of $\mathcal{BL}(\mathcal{M}_A,\mathcal{N}_B)$, we let $\rho_1,\rho_2\in \mathcal{BL}(\mathcal{M}_A,\mathcal{N}_B)$ and $0<t<1$.
We see from Theorem 2.1 that for $n=1,2$, there exists a PD $\{q^n_{k,j}:1\le k\le N_A,1\le j\le N_B\}$ such that
$${\rm{tr}}[(M_{a|x}\otimes N_{b|y})\rho_n]
=\sum_{k=1}^{N_A}\sum_{j=1}^{N_B}q^n_{k,j}\delta_{a,J_k(x)}\delta_{b,K_j(y)},\ \forall a,x,b,y.\eqno(2.10)$$
Thus, we get from Eq. (2.10) that $\forall a,x,b,y,$
\begin{eqnarray*}{\rm{tr}}[(M_{a|x}\otimes N_{b|y})(t\rho_1+(1-t)\rho_2]
&=&\sum_{k=1}^{N_A}\sum_{j=1}^{N_B}[tq^1_{k,j}+(1-t)q^2_{k,j}]\delta_{a,J_k(x)}\delta_{b,K_j(y)}\\
&=&\sum_{k=1}^{N_A}\sum_{j=1}^{N_B}tq_{k,j}\delta_{a,J_k(x)}\delta_{b,K_j(y)},
\end{eqnarray*}
where $q_{k,j}=tq^1_{k,j}+(1-t)q^2_{k,j}$ for all $k,j$. Clearly, $\sum_{k,j}q_{k,j}=1$. By using Theorem 2.1 again, we see that $t\rho_1+(1-t)\rho_2\in \mathcal{BL}(\mathcal{M}_A,\mathcal{N}_B)$.

Lastly, by using the fact that  $$\mathcal{BL}(AB)=\bigcap_{\mathcal{M}_A,\mathcal{N}_B}\mathcal{BL}(\mathcal{M}_A,\mathcal{N}_B),$$ we see that $\mathcal{BL}(AB)$ is a compact convex set. The proof is completed.

\section{Steerability of bipartite quantum states}

{\bf Definition 3.1.} (Steerability) Let $\rho^{AB}$ be a state of the system $AB$, and let $$\mathcal{M}_A=\{\{M_{a|x}\}_{a=1}^{o_A}:x=1,2,\ldots,m_A\}$$  be any  measurement assemblage of $A$.

(1) A state $\rho^{AB}$ of the system $AB$ is said to be {{\it unsteerable}} from $A$ to $B$ with respect to $\mathcal{M}_A$ if
there exists a PD $\{\pi_\lambda\}_{\lambda=1}^{d}$ and a set of states $\{\sigma_\lambda\}_{\lambda=1}^{d}\subset \mathcal{D}_B$ such that
$$\rho_{a|x}:={\rm{tr}}_A[(M_{a|x}\otimes 1_B)\rho^{AB}]=\sum_{\lambda=1}^d\pi_\lambda P_A({a|x},\lambda)\sigma_\lambda,\ \ \forall x,a,\eqno(3.1)$$
where $\{P_A({a|x},\lambda)\}_{a=1}^{o_A}$ is a PD for each $(a,x)$.
In this case, we also say that Eq. (3.1) is  an {\it LHS model} of $\rho^{AB}$ with respect to $\mathcal{M}_A$

(2) A state $\rho^{AB}$ is said to be {{\it steerable}}  from $A$ to $B$   with respect to  $\mathcal{M}_A$ if it is not unsteerable  from $A$ to $B$  with respect to  $\mathcal{M}_A$. In this case, we also say that $\rho^{AB}$ {\it exhibits quantum steering}   with respect to  $\mathcal{M}_A$.

 (3) A state $\rho^{AB}$ is said to be  {{\it unsteerable}}  from $A$ to $B$ if for any $\mathcal{M}_A$, $\rho^{AB}$ is {{\it unsteerable}}  from $A$ to $B$   with respect to  $\mathcal{M}_A$.

(4) A state  $\rho^{AB}$ is said to be {{\it steerable}}  from $A$ to $B$  if $\exists$ an $\mathcal{M}_A$ such that it is steerable  from $A$ to $B$ with respect to  $\mathcal{M}_A$, i.e. it is not unsteerable  from $A$ to $B$ with respect to  $\mathcal{M}_A$.

Symmetrically, we define unsteerability and steerability of a state from $B$ to $A$.

(5) A state  $\rho^{AB}$ is said to be is {{steerable}} if it is steerable from $A$ to $B$ or $B$ to $A$.

(6) A state  $\rho^{AB}$ is said to be {{unsteerable}} if it is not steerable, i.e. it is unsteerable both from $A$ to $B$, and  $B$ to $A$.

Here are some remarks to the definitions above.

{\bf Remark 3.1.} Denote by  ${\mathcal{US}}(A\rightarrow B, \mathcal{M}_A)$ the set of all states  which are unsteerable  from $A$ to $B$ with respect to  $\mathcal{M}_A$, by ${\mathcal{US}}(A\rightarrow B)$ the set of all states  which are ussteerable  from $A$ to $B$, and denote by  ${\mathcal{S}}(A\rightarrow B,\mathcal{M}_A)$ the set of all states  which are steerable  from $A$ to $B$ with respect to  $\mathcal{M}_A$, by ${\mathcal{S}}(A\vee B)$ the set of all states  which are steerable  from  either $A$ to $B$, or $B$ to $A$.
From definition above, we have
$${\mathcal{US}}(A\rightarrow B)=\bigcap_{\mathcal{M}_A}{\mathcal{US}}(A\rightarrow B, \mathcal{M}_A);$$
$${\mathcal{US}}(B\rightarrow A)=\bigcap_{\mathcal{M}_B}{\mathcal{US}}(B\rightarrow A, \mathcal{M}_B);$$
$${\mathcal{US}}(A\wedge B)={\mathcal{US}}(A\rightarrow B)\cap {\mathcal{US}}(B\rightarrow A);$$
$${\mathcal{S}}(A\vee B)={\mathcal{S}}(A\rightarrow B)\cup {\mathcal{S}}(B\rightarrow A).$$

{\bf Remark 3.2.} When  $\rho^{AB}\in {\mathcal{US}}(A\rightarrow B, \mathcal{M}_A)$, Eq. (3.1) holds. Thus, we have
$$\rho_B={\rm{tr}}_A\left[\sum_{a=1}^{o_A}(M_{a|x}\otimes 1_B)\rho^{AB}\right]=\sum_{\lambda=1}^d\pi_\lambda\sum_{a=1}^{o_A}P_A(a|x,\lambda)\sigma_\lambda.$$
Since $\sum_{a=1}^{o_A}P_A(a|x,\lambda)=1$ for all $\lambda$ and $x$, we get
$$\rho_B=\sum_{\lambda=1}^d\pi_\lambda\sigma_\lambda,\eqno(3.2)$$
which is independent of the choice of Alice's measurements $M^x$.  This means that the choice of Alice's measurements can not change (steer) Bob's state $\rho_B$, which is always given by Eq. (3.2).

Generally, the PD $\{\pi_\lambda\}_{\lambda=1}^{d}$ and the states $\{\sigma_\lambda\}_{\lambda=1}^{d}$ depend on the state $\rho^{AB}$ and the measurement assemblage $\mathcal{M}_A$.

The physical interpretation
is the following: when a state $\rho^{AB}$ is unsteerable with respect $\mathcal{M}_A$, Eq. (3.1) enables that Bob can interpret his conditional states $\rho_{a|x}:=\tr_A[(M_{a|x}\otimes 1_B)\rho^{AB}]$ as coming from the pre-existing states $\{\sigma_\lambda\}$ and the PD $\{\pi_\lambda\}$, where only the probabilities are changed due to the knowledge $\{P_A(a|x,\lambda)\}$ of Alice's measurement and result. Also, he can obtain his state $\rho_B$ from the pre-existing states $\{\sigma_\lambda\}$ and the PD $\{\pi_\lambda\}$ in light of Eq. (3.2).     Contrarily, when a state $\rho^{AB}$ is steerable with respect to $\mathcal{M}_A$, Bob must believe that Alice can remotely steer the
states in his lab by making measurements $\mathcal{M}_A$ on her side.

{\bf Example 3.1.} Let us now assume that Alice's measurements in  $\mathcal{M}_A$ are compatible,
in the sense of being jointly measurable \cite{CP}. This means
that there exists a single `parent' POV measurement $N=\{N_\lambda\}_{\lambda=1}^d$ such that
$\forall M^x=\{M_{a|x}\}_{a=1}^{o_A}\in \mathcal{M}_A$, there is $d$ PDs $\{P_A({a|x},\lambda)\}_{a=1}^{o_A}(\lambda=1,2,\ldots,d)$, such that $$M_{a|x}=\sum_{\lambda=1}^dP_A({a|x},\lambda)N_\lambda(a=1,2,\ldots,o_A).$$ Thus,  for any state $\rho^{AB}$ of the system $AB$, we have for each $(a,x)$,
$${\rm{tr}}_A[(M_{a|x}\otimes 1)\rho^{AB}]=\sum_{\lambda=1}^d P_A({a|x},\lambda){\rm{tr}}_A[(N_\lambda\otimes1)\rho^{AB}]
=\sum_{\lambda=1}^d\pi_\lambda P_A({a|x},\lambda)\sigma_\lambda,$$
where
$$\pi_\lambda={\rm{tr}}[(N_\lambda\otimes1)\rho^{AB}],\ \sigma_\lambda=\frac{1}{\pi_\lambda}{\rm{tr}}_A[(N_\lambda\otimes1)\rho^{AB}].$$
This shows that every state  $\rho^{AB}$ is unsteerable from $A$ to $B$ with respect to  a compatible measurement assemblage $\mathcal{M}_A$. Especially, when Alice has just one POV measurement $M=\{M_a\}_{a=1}^{o_A}$, i.e. $\mathcal{M}_A=\{M\}$, any state $\rho^{AB}$ of the system $AB$ is unsteerable from $A$ to $B$ with respect to  $\mathcal{M}_A$. Explicitly,
$${\rm{tr}}_A[(M_{a}\otimes 1)\rho^{AB}]=\sum_{\lambda=1}^{o_A}\pi_\lambda P_A({a}|M,\lambda)\sigma_\lambda,$$
where
$$\pi_\lambda={\rm{tr}}[(M_\lambda\otimes1)\rho^{AB}],\ P_A({a}|M,\lambda)=\delta_{\lambda,a}, \sigma_\lambda=\frac{1}{\pi_\lambda}{\rm{tr}}_A[(M_\lambda\otimes1)\rho^{AB}].$$

In a word, it is not possible that Alice wants to steer Bob with just one POVM.

{\bf Theorem 3.1.} {\it A state $\rho^{AB}$ of the system $AB$ is   {{unsteerable}} from $A$ to $B$ with respect to  $\mathcal{M}_A$ if and only if
there exists a PD $\{\pi_\lambda\}_{\lambda=1}^{d}$, a group of states $\{\sigma_\lambda\}_{\lambda=1}^{d}\subset \mathcal{D}_B$,  and  $dm_A$ PDs $\{P_A({a|x},\lambda)\}_{a=1}^{o_A}(1\le x\le m_A,1\le\lambda\le d)$  such that every local POVM $\{N_b\}_{b=1}^{o_B}$ of $B$, it holds that
$${\rm{tr}}[(M_{a|x}\otimes N_b)\rho^{AB}]=\sum_{\lambda=1}^d\pi_\lambda P_A({a|x},\lambda){\rm{tr}}(N_b\sigma_\lambda),\ \ \forall x,a,b. \eqno(3.3)$$
}

{\bf Proof.} {\it Necessity.} Let  $\rho^{AB}$ be  {{unsteerable}} from $A$ to $B$ with respect to  $\mathcal{M}_A$. Then by definition, there exists a PD $\{\pi_k\}_{k=1}^{d}$ and a group of states $\{\sigma_k\}_{k=1}^{d}\subset \mathcal{D}_B$  such that Eq. (3.1) holds for all $x,a.$  For any POVM $\{N_b\}_{b=1}^{o_B}$ of $B$, we see from Eq. (3.1) that $\forall x,a,b,$
\begin{eqnarray*}
{\rm{tr}}[(M_{a|x}\otimes N_b)\rho^{AB}]&=&{\rm{tr}}\left(N_b{\rm{tr}}_A[(M_{a|x}\otimes 1_B)\rho^{AB}]\right)\\
&=&{\rm{tr}}\left(\sum_{\lambda=1}^d\pi_\lambda P_A({a|x},\lambda)(N_b\sigma_\lambda)\right)\\
&=&\sum_{\lambda=1}^d\pi_\lambda P_A({a|x},\lambda){\rm{tr}}(N_b\sigma_\lambda).
\end{eqnarray*}

{\it Sufficiency.} Suppose that  Eq. (3.3) holds for every POVM $\{N_j\}_{j=1}^{o_B}$ of $B$.
Then for every $M^x=\{M_{a|x}\}_{a=1}^{o_A}\in \mathcal{M}_A$ and for  every projection $P$ on $\mathcal{H}_B$, using Eq. (3.3) for $N_1=P,N_2=I_B-P$ yields that for every $(x,a)$,
\begin{eqnarray*}
{\rm{tr}}\left(P{\rm{tr}}_A[(M_{a|x}\otimes 1_B)\rho^{AB}]\right)&=&
{\rm{tr}}[(M_{a|x}\otimes P)\rho^{AB}]\\
&=&\left(\sum_{\lambda=1}^d\pi_\lambda P_A(a|x,\lambda){\rm{tr}}(P\sigma_\lambda)\right)\\
&=&{\rm{tr}}\left(P\sum_{\lambda=1}^d\pi_\lambda P_A(a|x,\lambda)\sigma_\lambda\right).
\end{eqnarray*}
Thus,   for every $(x,a)$,
$$\left\<P,{\rm{tr}}_A[(M_{a|x}\otimes 1_B)\rho^{AB}]\right\>_{HS}=
\left\<P,\sum_{\lambda=1}^d\pi_\lambda P_A({a|x},\lambda)\sigma_\lambda\right\>_{HS},$$
where $\left\<X,Y\right\>_{HS}:= {\rm{tr}}(X^\dag Y)$ denotes the Hilbert-Schmidt inner product on the operator space $B(\mathcal{H}_B)$. Hence, for every $(x,a)$,
$${\rm{tr}}_A[(M_{a|x}\otimes 1_B)\rho^{AB}]=
\sum_{\lambda=1}^d\pi_\lambda P_A(a|x,\lambda)\sigma_\lambda.$$
This is just Eq.(3.1). Thus, $\rho^{AB}$ is  {{unsteerable}} from $A$ to $B$ with respect to  $\mathcal{M}_A$. The proof is completed.

We see from Corollary 3.1 that the steering of Alice to Bob needs to get a help from BoB.

Similarly, one can prove the following.

{\bf Theorem 3.2.} {\it A state $\rho^{AB}$ of the system $AB$ is   {{unsteerable}} from $A$ to $B$ if and only if for every $\mathcal{M}_A$,  there exists a PD $\{\pi_\lambda\}_{\lambda=1}^{d}$, a set of states $\{\sigma_\lambda\}_{\lambda=1}^{d}\subset \mathcal{D}_B$ and $dm_A$ PDs $\{P_A({a|x},\lambda)\}_{a=1}^{o_A}(1\le x\le m_A,1\le\lambda\le d)$ such that for every  POVM $\{N_b\}_{b=1}^{o_B}$ of $B$, it holds that
$${\rm{tr}}[(M_{a|x}\otimes N_b)\rho^{AB}]=\sum_{\lambda=1}^d\pi_\lambda P_A(a|x,k){\rm{tr}}(N_b\sigma_\lambda),\ \ \forall x,a,b,\eqno(3.4)$$}

Now, let us derive a very useful necessary and sufficient condition for a state to be unsteerable from $A$ to $B$ with respect to  $\mathcal{M}_A$. To do this, we consider the set $\Omega$ of all possible maps from $S_m=\{1,2,\ldots,m_A\}$ into $S_o=\{1,2,\ldots,o_A\}$. Clearly, $\Omega$ has just $N:=o_A^{m_A}$ elements and so can be written as
 $$\Omega=\{J_1,J_2,\ldots,J_N\}.$$
Each element $J$ of $\Omega$ denotes a ``measurement scenario", which assigns an outcome value $a$ for each POVM $x\equiv{M}^x$, that is, $J(x)=a$. We use $p(k,\lambda)$ to denote the probability of a measurement scenario $J_k$ to be used when Alice receives a classical message $\lambda$ in $\Lambda$, and $P(a,x,\lambda)$ to denote the probability of obtaining the outcome $a$ under the condition that Alice receives a classical message $\lambda$ in $\Lambda$ and chooses $\mathcal{M}^x$. Then the Law of Total Probability yields that
$$P(a,x,\lambda)=\sum_{k=1}^Np(k,\lambda)\delta_{a,J_k(x)},\ \ \forall a\in S_o,\forall x\in S_m,\eqno(3.5)$$
where
$$\sum_{k=1}^Np(k,\lambda)=1(\forall \lambda\in\Lambda),\ \ \sum_{a=1}^{o_A}\delta_{a,J_k(x)}=1(\forall k,x),\ \ \sum_{a=1}^{o_A}P(a,x,\lambda)=1(\forall a,x).$$
Please refer to \cite{Pusey} and \cite{CP} for Eq. (3.5).

Suppose that $\rho^{AB}$  is unsteerable from $A$ to $B$ with respect to  $\mathcal{M}_A$. Then by definition, there exists a PD $\{\pi_\lambda\}_{\lambda=1}^{d}$ and a set of states $\{\rho^B_\lambda\}_{\lambda=1}^{d}\subset \mathcal{D}_B$ such that
$${\rm{tr}}_A[(M_{a|x}\otimes 1_B)\rho^{AB}]=\sum_{\lambda=1}^d\pi_\lambda P_A({a|x},\lambda)\rho^B_\lambda,\ \ \forall a\in S_o,\forall x\in S_m.\eqno(3.6)$$
By taking traces of two sides, we get
$${\rm{tr}}[(M_{a|x}\otimes 1_B)\rho^{AB}]=\sum_{\lambda=1}^d\pi_\lambda P_A({a|x},\lambda),\ \ \forall a\in S_o,\forall x\in S_m.\eqno(3.7)$$
The left-hand side is the probability of having outcome $a$ when measurement $x$ is performed  and $P_A({a|x},\lambda)$ is the probability of obtaining the
outcome $a$ under the condition that Alice  receives a classical message $\lambda$ and chooses ${M}^x$. Thus, $P_A({a|x},\lambda)=P(a,x,\lambda)$ and so Eq. (3.5) yields that
$$P_A({a|x},\lambda)=\sum_{k=1}^Np(k,\lambda)\delta_{a,J_k(x)},\ \ \forall a\in S_o,\forall x\in S_m.$$
It follows from (3.6) that
$${\rm{tr}}_A[(M_{a|x}\otimes 1_B)\rho^{AB}]=\sum_{k=1}^N \delta_{a,J_k(x)}\sum_{\lambda=1}^d\pi_\lambda p(k,\lambda)\rho^B_\lambda,\ \ \forall a\in S_o,\forall x\in S_m.$$
By putting $\tau_k=\sum_{\lambda=1}^d\pi_\lambda p(k,\lambda)\rho^B_\lambda$, we obtain that
$${\rm{tr}}_A[(M_{a|x}\otimes 1_B)\rho^{AB}]=\sum_{k=1}^N\delta_{a,J_k(x)}\tau_k,\ \ \forall a\in S_o,\forall x\in S_m,\eqno(3.8)$$
satisfying $\tau_k\ge0$ for all $k$ and $\sum_{k=1}^N\tr(\tau_k)=1$. See \cite{CP,JM}.

Conversely, we suppose that there exists there exists positive operators  $\tau_k(k=1,2,\ldots,N)$ on $\mathcal{H}_B$ satisfying $\sum_{k=1}^N\tr(\tau_k)=1$ and such that (3.8) holds. Let  $\pi_k={{\rm{tr}}}(\tau_k),\sigma_k=\frac{1}{\pi_k}\tau_k$. Then (3.8) becomes
$${\rm{tr}}_A[(M_{a|x}\otimes 1)\rho^{AB}]=\sum_{k=1}^N\pi_k \delta_{a,J_k(x)}\sigma_k,\ \ \forall x\in S_m,\forall a\in S_o.\eqno(3.9)$$
Since $\sum_{a=1}^{o_A}\delta_{a,J_k(x)}=1$ for all $x\in S_m$ and all $k=1,2,\ldots,N$, by taking $P_A(a|x,k)=\delta_{a,J_k(x)}$ we see by definition that
$\rho^{AB}$  is unsteerable  from $A$ to $B$ with respect to  $\mathcal{M}_A$.

As a conclusion, we have established the following theorem.

{\bf Theorem 3.3.} {\it A state $\rho^{AB}$ of the system $AB$ is   {{unsteerable}} from $A$ to $B$ with respect to  $\mathcal{M}_A$ if and only if  there exists a family $\{\tau_k\}_{k=1}^{N}$ of positive operators $\tau_k$ on $\mathcal{H}_B$ with
$\sum_{k=1}^N\tr(\tau_k)=1$ such that (3.8) holds.}

It is remarkable to point out that positive operators $\tau_k$ in Eq. (3.8) depend only on the state $\rho^{AB}$ and are independent of the measurement operators $M_{a|x}$, while the deterministic PDs $\{\delta_{a,J_k(x)}\}_a$ depend only on the measurement operators $\{M_{a|x}\}$, independent of the state $\rho^{AB}$. Also, the number $N=o_A^{m_A}$ of terms of summation is fixed whenever the measurement assemblage $\mathcal{M}_A$ is given. This enables us to prove the following important properties of unsteerable states.

{\bf Corollary 3.1.} {\it ${\mathcal{US}}(A\rightarrow B,\mathcal{M}_A)$ is a compact convex subset of $\mathcal{D}_{AB}$.}

{\bf Proof.} Let $\rho_1,\rho_2\in {\mathcal{US}}(A\rightarrow B,\mathcal{M}_A)$ and $0<t<1$. Then by Theorem 3.3, there exist there exist families $\{\tau^1_k\}_{k=1}^{N}$ and $\{\tau^2_k\}_{k=1}^{N}$ of positive operators $\tau^i_k$ on ${\mathcal{H}}_B$ with
$\sum_{k=1}^N\tau^i_k)=1(i=1,2,k=1,2,\ldots,N)$ such that
$${\rm{tr}}_A[(M_{a|x}\otimes 1)\rho^{i}]=\sum_{k=1}^N\delta_{a,J_k(x)}\tau^i_k (i=1,2),\ \ \forall x\in S_m,\forall a\in S_o.$$
Thus, $\forall x\in S_m,\forall a\in S_o$, we have
$${\rm{tr}}_A[(M_{a|x}\otimes 1)(t\rho_{1}+(1-t)\rho_2]=\sum_{k=1}^N\delta_{a,J_k(x)}\tau_k,$$
where $\tau_k=t\tau^1_k+(1-t)\tau^2_k\ge0$ for all $k$ and $\sum_{k=1}^N\tr(\tau_k)=1.$ Thus, Theorem 3.3 implies that  $t\rho_{1}+(1-t)\rho_2$ is unsteerable from $A$ to $B$ with respect to  $\mathcal{M}_A$  and then ${\mathcal{US}}(A\rightarrow B,\mathcal{M}_A)$ is convex.

Next, let $\{\rho_m\}_{m=1}^\infty$ be a sequence in ${\mathcal{US}}(A\rightarrow B,\mathcal{M}_A)$ such that $\rho_m\rightarrow \rho$ as $m\rightarrow+\infty.$ By Theorem 3.3, there are positive operators $\tau^m_k(k=1,2,\ldots,N,m=1,2,\ldots)$ on $\mathcal{H}_B$ such that $\sum_{k=1}^N\tr(\tau^m_k)=1$  and
$${\rm{tr}}_A[(M_{a|x}\otimes 1_B)\rho_m]=\sum_{k=1}^N\delta_{a,J_k(x)}\tau^m_k, \ \ \forall n\in S_m, \forall n\in S_o,\eqno(3.10)$$
for all $m$. By the compactness of $\mathcal{D}_B$, we may assume that for each $k=1,2,\ldots,N$, $\{\tau^m_k\}_{m=1}^\infty$ is convergent and let $\tau^m_k\rightarrow \tau_k$ as  $m\rightarrow+\infty.$ Then by letting $m\rightarrow+\infty$ in (3.10), we get
$${\rm{tr}}_A[(M_{a|x}\otimes 1_B)\rho]=\sum_{k=1}^N\delta_{a,J_k(x)}\tau_k, \ \ \forall n\in S_m, \forall a\in S_o.\eqno(3.11)$$
Furthermore, since $\sum_{k=1}^N\tr(\tau^m_k)=1(m=1,2,\ldots)$, we see $\sum_{k=1}^N\tr(\tau_k)=1$. Clearly, $\tau_k\ge0$ for all $k$. Now, Theorem 3.3 shows that is unsteerable from $A$ to $B$ with respect to  $\mathcal{M}_A$  and therefore  ${\mathcal{US}}(A\rightarrow B,\mathcal{M}_A)$ is closed and then compact. The proof is completed.

{\bf Corollary 3.2.} {\it The set ${\mathcal{US}}(A\rightarrow B)$ is a compact convex subset of $\mathcal{D}_{AB}$ and ${\mathcal{S}}(A\rightarrow B)$ is open.}

 {\bf Proof.} From Remark 3.1, we know that
$${\mathcal{US}}(A\rightarrow B)=\bigcap_{\mathcal{M}_A}{\mathcal{US}}(A\rightarrow B, \mathcal{M}_A),$$
where the intersection was taken over all measurement assemblages $\mathcal{M}_A$ of $A$. It follows from Corollary 3.1 that ${\mathcal{US}}(A\rightarrow B)$ is compact and convex. The proof is completed.

As the end of this section, let us discuss some relationships among steerability, nonlocality,  entanglement and quantum correlations.
From Theorem 3.1 and Theorem 3.2, we see the following remarks.

(1) When $\rho^{AB}$ is unsteerable  from $A$ to $B$ with respect to  $\mathcal{M}_A$, we see from Definition 3.1 that
there exists a PD $\{\pi_\lambda\}_{\lambda=1}^{d}$, a set of states $\{\sigma_\lambda\}_{\lambda=1}^{d}\subset \mathcal{D}_B$,  and  $dm_A$ PDs $\{P_A({a|x},\lambda)\}_{a=1}^{o_A}(1\le x\le m_A,1\le\lambda\le d)$ such that
$${\rm{tr}}_A[(M_{a|x}\otimes 1_B)\rho^{AB}]=\sum_{\lambda=1}^d\pi_\lambda P_A({a|x},\lambda)\sigma_\lambda,\ \ \forall x,a.$$
Thus, for  any $\mathcal{M}_B$,
$${\rm{tr}}[(M_{a|x}\otimes N_{b|y})\rho^{AB}]
={\rm{tr}}(N_{b|y}{\rm{tr}}_A[(M_{a|x}\otimes 1_B)\rho^{AB}])\\
=\sum_k\pi_kP_A(a|x,k)P_B(b|y,k),$$
where $P_B(b|y,k)={\rm{tr}}(N_{b|y}\sigma_k).$
Thus, $\rho^{AB}$ is Bell local for $\mathcal{M}_A\otimes\mathcal{M}_B$.

(2) When $\rho^{AB}$ is unsteerable either from $A$ to $B$, or from $B$ to $A$, it is Bell local.
This shows that an unsteerable state must be Bell local, i.e. \begin{center}{${\mathcal{US}}(A\wedge B)={\mathcal{US}}(A\rightarrow B)\cap {\mathcal{US}}(B\rightarrow A)\subset {\mathcal{US}}(A\rightarrow B)\cup {\mathcal{US}}(B\rightarrow A)\subset \mathcal{BL}(AB).$}\end{center}

(3) When $\rho^{AB}=\sum_{k=1}^d\pi_k\rho^A_k\otimes\rho^B_k$ is separable, especially, classically-classically correlated \cite{Guo1,Guo2,Guo3}, we have for any $\mathcal{M}_A$,
$${\rm{tr}}_A[(M_{a|x}\otimes 1_B)\rho^{AB}]=\sum_{k=1}^d\pi_kP_A(a|x,k)\rho^B_k,\ \ \forall x,a,$$
where $P_A(a|x,k)={\rm{tr}}(M_{a|x}\rho^A_k).$ Thus, $\rho^{AB}$ is unsteerable from $A$ to $B$ with respect to  any $\mathcal{M}_A$.
Thus, $\rho^{AB}$ is unsteerable from $A$ to $B$ with respect to  any $\mathcal{M}_A$. Thus, $\rho^{AB}$ is unsteerable  from $A$ to $B$. Similarly, $\rho^{AB}$ is also unsteerable  from $B$ to $A$. A state which is steerable both from $A$ to $B$ and from $B$ to $A$ is said to be {\it two-way steerable}. A state which is steerable either from $A$ to $B$, or from $B$ to $A$ is said to be {\it one-way steerable}.

With the discussion above, we have the following relationships.
$${CC(AB)\supsetneq Sep(AB)\supsetneq {\mathcal{US}}(A\rightarrow B)\cap {\mathcal{US}}(B\rightarrow  A)\supsetneq {\mathcal{US}}(A\rightarrow  B)\cup {\mathcal{US}}(B\rightarrow A)\supsetneq {\mathcal{BL}}(AB),}$$
where $CC(AB)$ and $Sep(AB)$ are sets of all classically-classically (CC) correlated and separable states of $AB$, respectively.
Hence,
$${QC(AB)\supsetneq Ent(AB)\supsetneq {\mathcal{S}}(A\rightarrow B)\cup {\mathcal{S}}(B\rightarrow  A)\supsetneq {\mathcal{S}}(A\rightarrow  B)\cap {\mathcal{S}}(B\rightarrow A)\supset {\mathcal{BNL}}(AB),}$$
where

$QC(AB)=D(AB)\setminus CC(AB),$ the set of all quantum correlated states of $AB$;

$Ent(AB)=D(AB)\setminus Sep(AB),$ the set of all entangled states of $AB$;

${\mathcal{BNL}}(AB)=D(AB)\setminus {\mathcal{BL}}(AB)$, the set of all Bell nonlocal states of $AB$.

Consequently,
\begin{center}{{Bell locality $\Leftarrow$ Unsteerability  $\Leftarrow$ Separability  $\Leftarrow$ Classical correlation}},\end{center}
equivalently,
\begin{center}{{Bell nonlocality $\Rightarrow$ Steerability  $\Rightarrow$ Entanglement  $\Rightarrow$ Quantum correlation}}\end{center}

\section{A EPR-steering criteria}

{\bf Definition 4.1.} Two bases $e=\{|e_i\>\}_{i=1}^n$ and $f=\{|f_i\>\}_{i=1}^n$ for an $n$-dimensional Hilbert space $\mathcal{H}$ are said to be {\it disjoint} and denoted by $e\bigwedge f=0$ if $|e_i\>\<e_i|\ne|f_j\>\<f_j|$ for all $i,j$, equivalently,
$$(\C|e_i\>)\cap(\C|f_j\>)=\{0\},\ \ \forall i,j.\eqno(4.1)$$

Generally,
for every basis $e=\{|e_i\>\}_{i=1}^{n}$ for $\mathcal{H}$, if $U=[u_{ij}]$ is an $n\times n$ unitary matrix such that $|u_{ij}|<1$  for all $i,j$, then the bases $Ue:=\{\sum_{j=1}^nu_{ij}|e_j\>\}_{i=1}^{n}$ and $e$ are disjoint. Especially, if $\mathcal{F}_n$ is the $n$-order quantum Fourier transform, i.e.
$$\mathcal{F}_n=\frac{1}{\sqrt{n}}[\omega_n^{(k-1)(j-1)}]\ (\omega_n={\rm{e}}^{\frac{2\pi}{n}{\rm{i}}}),$$
whose $(k,j)$-entry is
$$u_{kj}=\frac{1}{\sqrt{n}}\omega_n^{(k-1)(j-1)}(k,j=1,2,\ldots,n),$$
 then $e$ and $\mathcal{F}_ne$ are disjoint.

{\bf Lemma 4.1.} {\it If $|x\>$ is a pure state in a Hilbert state $\mathcal{H} (\dim(\mathcal{H})\ge2)$ and $T$ is a bounded linear operator on $\mathcal{H}$ with $0\le T\le |x\>\<x|$, then $T=r|x\>\<x|$ for real number $0\le r\le 1$.}

{\bf Proof.} Put $M=\C|x\>,$ then $H=M\oplus M^\perp$. In this decomposition, we have
$$|x\>\<x|=\left(
             \begin{array}{cc}
               1 & 0 \\
               0 & 0 \\
             \end{array}
           \right),\ \ T=\left(
             \begin{array}{cc}
               r & 0 \\
               0 & 0 \\
             \end{array}
           \right)$$
since $\ker(|x\>\<x|)\subset\ker(T)$. Since $0\le T\le |x\>\<x|$, we have $0\le r\le1$. From these representations, we see that $T=r|x\>\<x|.$ The proof is completed.

{\bf Theorem 4.1.} {\it Let $\mathcal{M}_A$ be a set of POVMs on $A$ and $\rho^{AB}\in \mathcal{D}_{AB}$. Suppose that there exist two disjoint
 bases $e=\{|e_i\>\}_{i=1}^{d_B}$ and $f=\{|f_i\>\}_{i=1}^{d_B}$ for $\mathcal{H}_B$ and there are two POVMs $P=\{P_i:i=1,2,\ldots, d_B\}$ and $Q=\{Q_i:i=1,2,\ldots,d_B\}$ such that
 $$\tr_A((P_i\otimes 1_B)\rho^{AB})= c_i|e_i\>\<e_i| (i=1,2,\ldots,d_B),\eqno(4.2)$$
 $$\tr_A((Q_i\otimes 1_B)\rho^{AB})= d_i|f_i\>\<f_i| (i=1,2,\ldots,d_B),\eqno(4.3)$$
 with $c_id_i>0(i=1,2,\ldots,d_B)$. Then $\rho^{AB}$ is steerable from $A$ to $B$  with respect to  any $\mathcal{M}_A$ containing  POVMs $P$ and $Q$.}

 {\bf Proof.} In our setting, $m_A=o_A=d_B$.
Suppose that the state $\rho^{AB}$ is {{unsteerable}} from $A$ to $B$  with respect to  some $\mathcal{M}_A$ containing $P$ and $Q$. Then by definition, there exists a PD $\{\pi_k\}_{k=1}^d$ and a group of states $\{\sigma_k\}_{k=1}^d\subset \mathcal{D}_B$ such that for every $M=\{E_i\}_{i=1}^{m}$ in $\mathcal{M}_A$, it holds that
$${\rm{tr}}_A((E_i\otimes 1_B)\rho^{AB})=\sum_{k=1}^d\pi_{k}P_A(i|M,k)\sigma_k,\ \forall i=1,2, \ldots,m,\eqno{(4.4)}$$
where $P_A(i|M,k)\ge0$ with $\sum_{i=1}^{m}P_A(i|M,k)=1(k=1,2,\ldots,d).$ In this case,
$$\sum_{k=1}^d\pi_k\sigma_k=\rho_B.\eqno{(4.5)}$$
By using Eq. (4.4) for $P=\{P_i\}_{i=1}^{d_B}$ and $Q=\{Q_i\}_{i=1}^{d_B}$, respectively, and combining Eqs. (4.2) and (4.3), we obtain that
$$\sum_{k=1}^d\pi_{k}P_A(i|P,k)\sigma_k= c_i|e_i\>\<e_i|(i=1,2,\ldots,d_B),\eqno(4.6)$$
$$\sum_{k=1}^d\pi_{k}P_A(i|Q,k)\sigma_k= d_i|f_i\>\<f_i|(i=1,2,\ldots,d_B).\eqno(4.7)$$
From Eq. (4.6), we see that
$$0\le c_i^{-1}\pi_{k}P_A(i|P,k)\sigma_k\le|e_i\>\<e_i|(i=1,2,\ldots,d_B)$$
for each $k=1,2,\ldots,d$.
Therefore, Lemma 4.1, we know that for each $k=1,2,\ldots,d$ and each $i=1,2,\ldots,d_B$, there exists $a_{ik}\in [0,1]$ such that
$$c_i^{-1}\pi_{k}P_A(i|P,k)\sigma_k=a_{ki}|e_i\>\<e_i|.$$
Because that $\sum_{i=1}^{d_B}P_A(i|P,k)=1$ for all $k=1,2,\ldots,d,$  we conclude that for each $k$, there exists an $i_k$ such that   $P_A({i_k}|P,k)\ne0$ and so $$\pi_{k}\sigma_k=\frac{c_{i_k}a_{ki}}{P_A({i_k}|P,k)}|e_{i_k}\>\<e_{i_k}|.$$
 This shows that
$$\{\pi_1\sigma_1,\pi_2\sigma_2,\ldots,\pi_d\sigma_d\}\subset
\bigcup_{i=1}^{d_B}(\R|e_i\>\<e_i|):=S_P.$$
Similarly,
$$\{\pi_1\sigma_1,\pi_2\sigma_2,\ldots,\pi_d\sigma_d\}\subset
\bigcup_{i=1}^{d_B}(\R|f_i\>\<f_i|):=S_Q.$$
Thus, $\{\pi_1\sigma_1,\pi_2\sigma_2,\ldots,\pi_d\sigma_d\}\subset S_P\bigcap S_Q.$ Since $e$ and $f$ are disjoint, $S_P\bigcap S_Q=\{0\}$ and so $\pi_k\sigma_k=0$ for all $k=1,2,\ldots,d$.  This contradicts Eq. (4.5). The proof is completed.

{\bf Corollary 4.1.} {\it Let $\{|\varepsilon_i\>\}_{i=1}^n$ be a real orthonormal basis for $\H_A=\H_B=\C^n$ and $|\psi\>=\frac{1}{\sqrt{n}}\sum_{i=1}^n|\varepsilon_i\>|\varepsilon_i\>$.  Then $\rho^{AB}=|\psi\>\<\psi|$ is steerable from $A$ to $B$  with respect to  any $\mathcal{M}_A$ containing POVMs $P$ and $Q$, in which $P=\{|e_i\>\<e_i|\}_{i=1}^n$ and $Q=\{|f_j\>\<f_j|\}_{j=1}^n$   where $e=\{|e_i\>\}_{i=1}^n$ is any basis for $\H_A=\C^n$ and $f=\mathcal{F}_ne=\{|f_i\>\}_{i=1}^n$.}

{\bf Proof.} First we compute that
$$\tr_A[(|x^*\>\<x^*|\otimes I_B)\rho^{AB}]=\frac{1}{n}|x\>\<x|, \ \ \forall |x\>\in \C^n,$$
where $|x^*\>$ denotes the conjugation of $|x\>$.
Since $f$ and $e$ are disjoint bases and
$$\tr_A((|e_i^*\>\<e_i^*|\otimes I_B)\rho^{AB})=\frac{1}{n}|e_i\>\<e_i|(i=1,2,\ldots,n),$$
$$\tr_A((|f_i^*\>\<f_i^*|\otimes I_B)\rho^{AB})=\frac{1}{n}|f_i\>\<f_i|(i=1,2,\ldots,n),$$
we see from Theorem 4.1 that $\rho^{AB}=|\psi\>\<\psi|$ is steerable from $A$ to $B$ with respect to  any $\mathcal{M}_A$ containing POVMs $P$ and $Q$. The proof is completed.

{\bf Example 4.1.} The bipartite maximally entangled state $$|\psi\>_{AB}=\frac{1}{\sqrt{2}}\left(|00\>+|11\>\right),$$
i.e. $\rho^{AB}=|\psi\>\<\psi|$,
is steerable from $A$ to $B$ with respect to  any $\mathcal{M}_A$ containing POVMs $\{|0\>\<0|,|1\>\<1|\}$ and $\{|f_1\>\<f_1|,|f_2\>\<f_2|\}$ where
$$|f_1\>=\frac{1}{\sqrt{2}}(1,1)^T, |f_2\>=\frac{1}{\sqrt{2}}(1,-1)^T.$$

{\bf Proof.} Use Corollary 4.1 for $|\varepsilon_1\>=|e_1\>=|0\>,|\varepsilon_2\>=|e_2\>=|1\>$. The proof is completed.

The following result shows that steerability is invariant under a local unitary transformation.

{\bf Theorem 4.2.} {\it Let $\rho\in \mathcal{D}_{AB}$ and let $U:\H_A\rightarrow \mathcal{K}_A$ and $V:\H_B\rightarrow \mathcal{K}_B$ be unitary operators, $\rho'=(U\otimes V)\rho(U^\dag\otimes V^\dag)$, and let $\mathcal{M}=\{M^x:x=1,2,\ldots,m_A\}$ be a set of POVMs (resp. projection measurements) of system $\H_A$. Then

(1) $\rho'\in D(\mathcal{K}_A\otimes \mathcal{K}_B)$.

(2)  $U\mathcal{M}U^\dag:=\{UM^xU^\dag:x=1,2,\ldots,m_A\}$ is a set of  POVMs  (resp. resp. projection measurements) of system $\mathcal{K}_A$ where $UM^xU^\dag=\{UM_{a|x}U^\dag:a=1,2,\ldots,o_A\}$ if $M^x=\{M_{a|x}:a=1,2,\ldots,o_A\}$.

(3) $\rho$ is unsteerable from $A$ to $B$ with $\mathcal{M}$ if and only if $\rho'$ is unsteerable from $A$ to $B$ with $U\mathcal{M}U^\dag$.

(4) $\rho$ is unsteerable from $A$ to $B$ if and only if $\rho'$ is unsteerable from $A$ to $B$.

(5) $\rho$ is unsteerable if and only if $\rho'$ is unsteerable.

(6) $\rho$ is steerable if and only if $\rho'$ is steerable.}

{\bf Proof.} (1) Denote $\rho'=(U\otimes V)\rho(U^\dag\otimes V^\dag)$. For any $|x\>\in K_A\otimes K_B$, by writing $|y\>=(U^\dag\otimes V^\dag)|x\>=|(U^\dag\otimes V^\dag)x\>$ we have
$$\<x|\rho'|x\>=\<(U^\dag\otimes V^\dag)x|\rho|(U^\dag\otimes V^\dag)x\>=\<y|\rho|y\>\geq 0,$$
and so $\rho'\geq0$. For any orthonormal basis $\{|x_i\>\}$ for $\mathcal{K}_A\otimes \mathcal{K}_B$, we have $\{(U^\dag\otimes V^\dag)|x_i\>\}$ is an orthonormal basis for $H_A\otimes H_B$ and so $$\tr\rho'=\sum_i\<x_i|(U\otimes V)\rho(U^\dag\otimes V^\dag)|x_i\>=\sum_i\<(U^\dag\otimes V^\dag) x_i|\rho|(U^\dag\otimes V^\dag)x_i\>=1.$$
Thus, $\rho'\in D(\mathcal{K}_A\otimes \mathcal{K}_B)$.

(2) Clearly.

(3) Suppose that $\rho$ is unsteerable from $A$ to $B$ with respect to  $\mathcal{M}$, then there exists a PD $\{\pi_k\}$ and states $\sigma_k\in \mathcal{D}_B$ such that
$${\rm{tr}}_A((M_{a|x}\otimes I_B)\rho)=\sum_k\pi_kP_A(M_{a|x},k)\sigma_k,\ \ \forall a,x\eqno(4.8)$$
for some $P_A(M_{a|x},k)\ge0$ with $\sum_{a}P_A(M_{a|x},k)=1.$ When $\rho=C\otimes D$, we  compute that $\rho'=UCU^\dag\otimes VDV^\dag$ and so for every operator $T$ on $\mathcal{H}_A,$
\begin{eqnarray*}
{\rm{tr}}_A[(UTU^\dag\otimes I_B)\rho']&=&
{\rm{tr}}_A(UTCU^\dag\otimes VDV^\dag)\\
&=&\tr(TC)\cdot VDV^\dag\\
&=&V\cdot \tr_A[(T\otimes I_B)(C\otimes D)]\cdot V^\dag\\
&=&V\cdot \tr_A[(T\otimes I_B)\rho]\cdot V^\dag.
\end{eqnarray*}
Generally, by writing $\rho=\sum_jC_j\otimes D_j$ we get that
$${\rm{tr}}_A[(UTU^\dag\otimes I_B)\rho']=V\cdot \tr_A[(T\otimes I_B)\rho]\cdot V^\dag.$$
By using this identity for $T=M_{a|x}$ and Eq. (4.8), we see  that
$${\rm{tr}}_A((UM_{a|x}U^\dag\otimes I_B)\rho')=\sum_k\pi_kP_A(M_{a|x},k)V\sigma_kV^\dag,\ \  \forall a,x.$$
Since $\{V\sigma_kV^\dag\}\subset D(K_B)$, we conclude that $\rho'$ is unsteerable from $A$ to $B$ with respect to  $UMU^\dag$.  By using this conclusion, we see that if  $\rho'$ is unsteerable from $A$ to $B$ with $UM^xU^\dag$, then $\rho$ is unsteerable from $A$ to $B$ with respect to  $U^\dag(M^xU^\dag)U=\mathcal{M}$.

(4)-(6): Use (1)-(3). The proof is completed.

{\bf Corollary 4.2.} {\it Suppose that $\{|\theta_i\>\}_{i=1}^{n}$ and $\{|\eta_i\>\}_{i=1}^{n}$ are bases for $\H_A=\H_B=\C^n$, $|\varphi\>=\frac{1}{\sqrt{n}}\sum_{i=1}^n|\theta_i\>|\eta_i\>$.
For any basis $e=\{|e_i\>\}_{i=1}^n$ for $\C^n$, let  $P=\{|e_i\>\<e_i|\}_{i=1}^n, Q=\{|f_j\>\<f_j|\}_{j=1}^n$ where $f=\mathcal{F}_ne=\{|f_i\>\}_{i=1}^n$. Then $\rho^{AB}=|\varphi\>\<\varphi|$ is steerable from $A$ to $B$  with respect to  any $\mathcal{M}_A$ containing $U^\dag PU$ and $U^\dag QU$ where $U$ is the  unitary operator on $\C^n$ satisfying
$U|\theta_i\>=|\varepsilon_i\>(\forall i)$
and $\{|\varepsilon_i\>\}_{i=1}^n$ is a real ONB for $\C^n$.}

{\bf Proof.} Let $V$ be the unitary operator on $\C^n$ such that
$V|\eta_i\>=|\varepsilon_i\>$ for all $i=1,2,\ldots,n$.
Since $$(U\otimes V)|\varphi\>=\frac{1}{\sqrt{n}}\sum_{i=1}^n|\varepsilon_i\>|\varepsilon_i\>:=|\psi\>,$$
 $\rho':=(U\otimes V)\rho(U^\dag\otimes V^\dag)=|\psi\>\<\psi|,$ which is steerable from $A$ to $B$ with respect to  any $\mathcal{M}_A$ containing $P$ and $Q$ (Corollary 4.1). Therefore, Theorem 4.2 yields that $\rho$ is steerable from $A$ to $B$ with respect to  any $\mathcal{M}_A$ containing $U^\dag PU$ and $U^\dag QU$. The proof is completed.

{\bf Example 4.2.} The bipartite maximally entangled state $$|\psi\>_{AB}=\frac{1}{\sqrt{2}}\left(|01\>+|10\>\right),$$
i.e. $\rho^{AB}=|\psi\>\<\psi|$
is steerable from $A$ to $B$
with respect to any $\mathcal{M}_A$ containing $\{|0\>\<0|,|1\>\<1|\}$ and $\{|f_1\>\<f_1|,|f_2\>\<f_2|\}$ where
$$|f_1\>=\frac{1}{\sqrt{2}}(1,1)^T, |f_2\>=\frac{1}{\sqrt{2}}(1,-1)^T.$$

{\bf Proof.} Use Corollary 4.2 for $|\theta_1\>=|e_1\>=|\varepsilon_1\>=|0\>,|\theta_2\>=|e_2\>=|\varepsilon_2\>=|1\>$,   $|\eta_1\>=|1\>,|\eta_2\>=|0\>, U=I.$ The proof is completed.

{\bf Lemma 4.2.} {\it  Let $\varepsilon=\{|i\>\}_{i=1}^n$ be the canonical $0-1$ basis for $\C^n$,  $|\psi\>=\sum_{i=1}^r\mu_i|i\>|i\>$ with $1<r\le n$ be an entangled pure state of $\C^n\otimes\C^n$.  Put $f=\mathcal{F}_n\varepsilon=\{|f_i\>\}_{i=1}^n$,  $P=\{|i\>\<i|\}_{i=1}^n$ and $Q=\{|f_j\>\<f_j|\}_{j=1}^n$.
Then  $\rho^{AB}=|\psi\>\<\psi|$ is steerable from $A$ to $B$  with respect to any $\mathcal{M}_A$ containing POVMs $P$ and $Q$.}

 {\bf Proof.}  We compute that
 $$\tr_A[(|x\>\<x|\otimes I_B)\rho^{AB}]=|x^\star\>\<x^\star|, \ \ \forall |x\>=\sum_{k=1}^n{a_k}|k\>\in \C^n,\eqno(4.9)$$
where $|x^\star\>=\sum_{k=1}^r\mu_k{a_k}^*|k\>$.  Especially,
$$\tr_A[(|i\>\<i|\otimes I_B)\rho^{AB}]=|i^\star\>\<i^\star|,\ \ \tr_A[(|f_j\>\<f_j|\otimes I_B)\rho^{AB}]=|f_j^\star\>\<f_j^\star|,\eqno(4.10)$$
for all $i,j=1,2,\ldots,n$.

Suppose that the state $\rho^{AB}$ is {{unsteerable}} from $A$ to $B$  with respect to  some $\mathcal{M}_A$ containing $P$ and $Q$. Then by definition, there exists a PD $\{\pi_k\}_{k=1}^d$ with positive probabilities   and a group of states $\{\sigma_k\}_{k=1}^d\subset \mathcal{D}_B$ such that for every $M=\{E_i\}_{i=1}^{m}$ in $\mathcal{M}_A$, it holds that
$${\rm{tr}}_A[(E_i\otimes 1_B)\rho^{AB}]=\sum_{k=1}^d\pi_{k}P_A(i|M,k)\sigma_k,\ \forall i=1,2, \ldots,m,\eqno{(4.11)}$$
where $P_A(i|M,k)\ge0$ with $\sum_{i=1}^{m}P_A(i|M,k)=1(k=1,2,\ldots,d).$ In this case,
$$\sum_{k=1}^d\pi_k\sigma_k=\rho_B.\eqno{(4.12)}$$
By using Eq. (4.11) for $P$ and $Q$, respectively, and combining Eq. (4.10), we obtain that
$$\sum_{k=1}^d\pi_{k}P_A(i|P,k)\sigma_k= |i^\star\>\<i^\star|(i=1,2,\ldots,n),\eqno(4.13)$$
$$\sum_{k=1}^d\pi_{k}P_A(j|Q,k)\sigma_k= |f_j^\star\>\<f_j^\star|(j=1,2,\ldots,n).\eqno(4.14)$$
Clearly,
$$|i^\star\>=\mu_i|i\>(1\le i\le r), |i^\star\>=0(r< i\le n); |f_j^\star\>=\sum_{k=1}^r\mu_kb_k^{(j)}|k\>(1\le j\le n),$$
where $b_k^{(j)}=\<j|\mathcal{F}_n|k\>$ satisfy $|f_j\>=\sum_{k=1}^nb_k^{(j)}|k\>.$ From the structure of The Fourier transformation $\mathcal{F}_n$, we know that each $b_k^{(j)}$ is not zero.

From Eq. (4.14) and Lemma 4.1, we know that for each $k=1,2,\ldots,d$ and each $i=1,2,\ldots,r$, there exists $a_{ik}\ne 0$ such that
$$\pi_{k}P_A(i|P,k)\sigma_k=a_{ik}|i^\star\>\<i^\star|.$$
Because that $\sum_{i=1}^{r}P_A(i|P,k)=1$ for all $k=1,2,\ldots,d,$  we conclude that for each $k$, there exists an $1\le i_k\le r$ such that   $P_A({i_k}|P,k)>0$ and so $$\pi_{k}\sigma_k=
\frac{a_{ki}}{P_A({i_k}|P,k)}|\varepsilon_{i_k}^\star\>\<\varepsilon_{i_k}^\star|.$$

Similarly,  for each $k$, there exists an $1\le j_k\le n$ such that   $P_A({j_k}|Q,k)>0$ and
so $$\pi_{k}\sigma_k=
\frac{b_{ki}}{P_A({j_k}|Q,k)}|f_{j_k}^\star\>\<f_{j_k}^\star|,$$
where $b_{ki}\ne0$. Thus, $|\varepsilon_{i_k}^\star\>=c_k|f_{j_k}^\star\>(k=1,2,\ldots,d)$ for some nonzero constants $c_k$, that is,
$$\mu_{i_k}|\varepsilon_{i_k}\>=c_k\sum_{m=1}^r\mu_mb_m^{(j_k)}|\varepsilon_{m}\>.$$
This shows that $b_m^{(j_k)}=0$ for all $m\ne i_k.$ Since $r>1$, such an $m$ does exist. This contradicts the fact that $b_k^{(j)}\ne0$ for all $k,j$.  Therefore,
 $\rho^{AB}=|\psi\>\<\psi|$ is steerable from $A$ to $B$ with respect to  any $\mathcal{M}_A$ containing POVMs $P$ and $Q$. The proof is completed.

{\bf Theorem 4.3.} {\it  Let $|\psi\> $ be an entangled pure state of $\C^n\otimes\C^n$.  Then there exist two POVMs $P$ and $Q$ such that  $\rho^{AB}=|\psi\>\<\psi|$ is steerable from $A$ to $B$  with respect to any $\mathcal{M}_A$ containing POVMs $P$ and $Q$.}

 {\bf Proof.}  Since $|\psi\> $ is  an entangled pure state of $\C^n\otimes\C^n$, it has Schmidt decomposition $|\psi\>=\sum_{i=1}^r\mu_i|\varepsilon_i\>|\eta_i\>$ where  $\{|\varepsilon_i\>\}_{i=1}^n$ and  $\{|\eta_i\>\}_{i=1}^n$ are orthonormal bases for $\C^n$ and $\mu_i>0$ for all $i=1,2,\ldots,r$ with $1<r\le n$.  Choose unitary operators $U$ and $V$ on $\C^n$ such that
 $$|\varepsilon_i\>=U|i\>, |\eta_i\>=V|i\>(i=1,2,\ldots,n)$$
  where $\{|i\>\}_{i=1}^n$ is the canonical $0-1$ basis for $\C^n$. Since $(U^\dag\otimes V^\dag)|\psi\>=\sum_{i=1}^r\mu_i|i\>|i\>$, it follows from Lemma 4.2 that $(U^\dag\otimes V^\dag)\rho^{AB}(U\otimes V)$ is steerable from $A$ to $B$ with any   $\mathcal{M}_A$ containing POVMs $\{|i\>\<i|\}_{i=1}^n$ and $\{\mathcal{F}_n|j\>\<j|\mathcal{F}_n^\dag\}_{j=1}^n$. By using Theorem 4.2, we know that the state $\rho^{AB}$ is steerable from $A$ to $B$ with any   $\mathcal{M}_A$ containing POVMs $P=\{U|i\>\<i|U^\dag\}_{i=1}^n$ and $Q=\{U\mathcal{F}_n|j\>\<j|\mathcal{F}_n^\dag U^\dag\}_{j=1}^n$. The proof is completed.

\section{Conclusions}

In this note, we have obtained some characterizations of Bell locality and EPR steerability of bipartite states and proved that the set of all Bell local states and the set of all unsteerable states are both convex and compact. The compactness of these sets are useful for quantifying Bell locality and EPR steerability. From the convexity of ${\mathcal{US}}(A\rightarrow B,\mathcal{M}_A)$, we see that when a mixed state with spectral decomposition $\rho=\sum_i\lambda_i|\psi_i\>\<\psi_i|$ is steerable from $A$ to $B$ with respect to  $\mathcal{M}_A$, there exists an $i$ such that  $|\psi_i\>\<\psi_i|$ is steerable from $A$ to $B$ with respect to  $\mathcal{M}_A$. From the convexity of ${\mathcal{US}}(A\rightarrow B)$, we see that when a mixed state with spectral decomposition $\rho=\sum_i\lambda_i|\psi_i\>\<\psi_i|$ is steerable from $A$ to $B$, there exists an $\mathcal{M}_A$ and an $i$ such that  $|\psi_i\>\<\psi_i|$ is steerable from $A$ to $B$ with respect to  $\mathcal{M}_A$. Since ${\mathcal{S}}(A\rightarrow B)$ is open, we conclude that when a state $\rho^{AB}$ is steerable from $A$ to $B$, all states close to $\rho^{AB}$ are steerable from $A$ to $B$.

We have also proved that any locally unitary operation do not change steerability. By using this fact and  proving a EPR-steering criteria, we prove that  any maximally entangled pure state of $\C^n\otimes \C^n$ is steerable from $A$ to $B$ with respect to  two projection measurements.

Moreover, convexity and compactness of ${\mathcal{BL}}(AB)$ implies that for every Bell nonlocal state $\sigma^{AB}$, there exists a Hermitian operator $\mathcal{L}$ on $\H_A\otimes\H_B$ such that
$$\tr(\mathcal{L}\rho^{AB})\ge0(\forall \rho^{AB}\in{\mathcal{BL}}(AB)){\mbox{\ and\ }} \tr(\mathcal{L}\sigma^{AB})<0.$$
Such an $\mathcal{L}$ is said to be a {{Bell nonlocality witness}}.
The steerability witness can be defined similarly.

\end{document}